\documentclass[copyright]{eptcs}

\usepackage{amsfonts, amsmath, amssymb, xcolor}
\usepackage{graphicx}

\newcommand{\Cu}{C^{\ast}} 
\newcommand{\Tr}{\mbox{Tr}}
\newcommand{\tr}{\mbox{tr}}
\newcommand{\R}{\mathbb{R}}

\newcommand{\M}{\mathbf{M}}
\newcommand{\N}{\mathbf{N}}
\renewcommand{\O}{\mathbb{O}}
\newcommand{\C}{\mathbb{C}}
\newcommand{\Q}{\mathbb{H}}
\newcommand{\K}{\mathbb{K}}
\newcommand{\Cat}{\mathcal{C}} 
\renewcommand{\H}{\mathbf{H}}

\newcommand{\IP}{\langle \  | \ \rangle}

\newcommand{\CJP}{\mbox{\bf CJP}}
\newcommand{\URSE}{\mbox{\bf URSE}}
\newcommand{\URUE}{\mbox{\bf URUE}}
\newcommand{\RSE}{\mbox{\bf RSE}}
\newcommand{\CQM}{{\mathbb C}{\bf QM}}
\newcommand{\EJC}{\mbox{\bf EJC}}
\newcommand{\stalg}{\mbox{\bf $\ast$-ALG}} 
\renewcommand{\bar}{\overline}
\renewcommand{\hat}{\widehat}
\newcommand{\hotimes}{\widetilde{\otimes}}

\newcommand{\id}{\mbox{id}}
\newcommand{\1}{\mathbf{1}}
\newcommand{\sa}{\mbox{\small sa}}

\newcommand{\ex}{\mbox{\em ex}}
\renewcommand{\sp}{\mbox{\em sp}}

\newcommand{\op}{\text{op}}

\newcommand{\tempout}[1]{{}}
\newcommand*{\jProd}{\raisebox{-0.88ex}{\scalebox{2.5}{$\cdot$}}}

\providecommand{\event}{QPL 2015} 
\title{Some Nearly Quantum Theories}
\author{Howard Barnum
\institute{University of New Mexico\\ Albuquerque, USA}
\email{hnbarnum@aol.com}
\and
Matthew A.\ Graydon
\institute{Perimeter Institute for Theoretical Physics\\
Waterloo, Canada}
\institute{Institute for Quantum Computing, University of Waterloo\\
Waterloo, Canada}
\email{mgraydon@perimeterinstitute.ca}
\and
Alexander Wilce
\institute{Susquehanna University\\
Selinsgrove, USA}
\email{wilce@susqu.edu}
}
\def\titlerunning{Some Nearly Quantum Theories}
\def\authorrunning{H.\ Barnum, M.~A.\ Graydon \& A.\ Wilce}\begin{document}
\maketitle

\begin{abstract}
We consider possible non-signaling composites of probabilistic models
based on euclidean Jordan algebras.  Subject to some reasonable
constraints, we show that no such composite exists having the
exceptional Jordan algebra as a direct summand. We then construct
several dagger compact categories of such Jordan-algebraic models. One
of these neatly unifies real, complex and quaternionic mixed-state
quantum mechanics, with the exception of the
quaternionic ``bit".  Another is similar, except in that (i) it
excludes the quaternionic bit, and (ii) the composite of two complex
quantum systems comes with an extra classical bit.  In both of these
categories, states are morphisms from systems to the tensor unit,
which helps give the categorical structure a clear operational interpretation.
A no-go result shows that the first
of these categories, at least,  
cannot be extended to include spin factors other than the (real, complex,
  and quaternionic) quantum bits, while preserving the representation
of states as morphisms.  The same is true for attempts to extend the 
second category to even-dimensional spin-factors. 
Interesting phenomena exhibited by
  some composites in these categories include failure of local
  tomography, supermultiplicativity of the maximal number of mutually
  distinguishable states, and mixed states whose marginals are pure.
\end{abstract}

\section{Introduction}
A series of recent papers \cite{Wilce09, MU, Wilce11, Wilce12, BMU}
have shown that any of various packages of probabilistic or
information-theoretic axioms force the state spaces of a finite-dimensional 
probabilistic theory to be those of formally real, or euclidean, Jordan
algebras. Thus, euclidean Jordan algebras (hereafter, EJAs) form a
natural class of probabilistic models. Moreover, it is one that
keeps us in the general neighborhood of standard quantum
mechanics, owing to the classification of simple EJAs as self-adjoint
parts of real, complex and quaternionic matrix algebras (corresponding
to real, complex and quaternionic quantum systems), the exceptional
Jordan algebra of self-adjoint $3 \times 3$ matrices over the
octonions, and one further class, the so-called {\em spin factors}.
The latter are essentially ``bits": their state-spaces are balls of
arbitrary dimension, with pairs of antipodal points 
representing maximal sets of sharply
distinguishable states.\footnote{Where this ball has dimension $2$, $3$ or $5$, 
these are just the state spaces of real, complex and quaternionic 
quantum bits.}\\[0.2cm]
\noindent This raises the question of whether one can construct probabilistic
{\em theories} (as opposed to a collection of models of individual
systems) in which finite-dimensional complex quantum systems can be accommodated together
with several --- perhaps all --- of the other basic types of EJAs listed
above. Ideally, these would be symmetric monoidal categories; even
better, we might hope to obtain compact closed, or still better,
dagger-compact, categories of EJAs \cite{Abramsky-Coecke}. Also, one would like the resulting
theory to embrace mixed states and CP mappings.\\[0.2cm]
\noindent In this paper, we exhibit two dagger-compact categories of EJAs ---
called $\URUE$ and $\URSE$, acronyms that will be explained below
--- that include all real, complex and quaternionic matrix algebras,
with one exception: the quaternionic bit, or ``quabit'', represented by $M_2(\Q)_{\sa}$, cannot be added to $\URUE$ without destroying compact closure and the representation of states as morphisms. $\URSE$ includes a faithful copy of finite-dimensional 
complex quantum mechanics, while in $\URUE$, composites of complex quantum systems come with an extra
classical bit --- that is, a $\{0,1\}$ valued superselection rule.\\[0.2cm]
\noindent There is scant hope of including more exotic Jordan
algebras in a satisfactory categorical scheme. Even allowing for a
very liberal definition of composite (our Definition 1 below), the
exceptional Jordan algebra is ruled out altogether (Corollary 1), while non-quantum spin
factors are ruled out if we want to regard states as morphisms --- in
particular, if we demand compact closure (see Example 1). Combined with the results
of (any of) the papers cited above that derive a euclidean
Jordan-algebraic structure from information-theoretic assumptions,
these results provide a compelling motivation for a kind of unified
quantum theory that accommodates real, complex and quaternionic
quantum systems (possibly modulo the quabit)  and permits the
formation of composites of these.\\[0.2cm]
\noindent A condition frequently invoked to rule out real and
quaternionic QM is {\em local tomography}: the doctrine that the state
of a composite of two systems should depend only on the joint
probabilities it assigns to measurement outcomes on the component
systems. Indeed, it can be shown \cite{BW12} that standard complex QM
with superselection rules is the only dagger-compact category of EJAs
that includes the qubit. Accordingly, $\URUE$ and $\URSE$ are {\em
  not} locally tomographic.  In our view, the very existence of these
quite reasonable, well-behaved categories suggests that local
tomography is not as well-motivated as is sometimes supposed.\\[0.2cm]
A broadly similar proposal is advanced by Baez \cite{Baez}, who points out that one can view real and quaternionic quantum systems as pairs $(\H,J)$, where $\H$ is a complex Hilbert space and $J$ is an anti-unitary 
satisfying $J^2 = \1$ (the real case) or $J^2 = -\1$ (the quaternionic case). This yields a symmetric monoidal 
category in which objects are such pairs, morphisms $(\H_1, J_1) \rightarrow (\H_2, J_2)$ are 
linear mappings intertwining $J_1$ and $J_2$, and $(\H_1, J_1) \otimes (\H_2, J_2) = (\H_1 \otimes \H_2, J_1 \otimes J_2)$. The precise connection between this approach and ours is still under study.\\[0.2cm]
A forthcoming paper \cite{BGW} will provide more details, including full proofs of the results presented here, as well as additional results obtained since our presentation of this paper at QPL XII.  
 
\section{Euclidean Jordan algebras} 

We begin with a concise review of some basic Jordan-algebraic background. References for this section 
are \cite{Alfsen-Shultz} and \cite{FK}. A {\em euclidean Jordan algebra} (hereafter: EJA) is a finite-dimensional commutative real algebra $(A,\jProd)$ with a multiplicative unit element $u$, 
satisfying the {\em Jordan identity} 
\[a^2 \jProd (a \jProd b) = a \jProd (a^2 \jProd b)\]
for all $a,b \in A$, and  
equipped with an inner product satisfying 
\[\langle a \jProd b | c \rangle = \langle b | a \jProd c \rangle\] 
for all $a,b,c \in A$.  The basic example is the self-adjoint part $M_{\sa}$ of a real, complex or quaternionic 
matrix algebra $M$, with 
$a \jProd b = (ab + ba)/2$ and with $\langle a | b \rangle = \tr(ab)$.  
Any Jordan subalgebra of an EJA is 
also an EJA. So, too, is the {\em spin factor} 
$V_n = \R \times \R^n$, with the obvious inner product and with 
\[(t,x) \jProd (s,y) = (ts + \langle x | y \rangle, ty + sx):\]
this can be embedded (universally) in $M_{2^{k}}(\C)_{\sa}$ for $n=2k$ and $M_{2^{k}}(\C)_{\sa}\oplus M_{2^{k}}(\C)_{\sa}$ for $n=2k+1$. Moreover, one can show that 
\[V_2 \simeq M_2(\R)_{\sa}, \ V_3 \cong M_{2}(\C)_{\sa}, \ \mbox{and} \ V_5 \simeq M_{2}(\Q)_{\sa}.\]\\[0.2cm]
\noindent{\bf Classification} Direct sums of EJAs are also EJAs, so we can obtain more examples by forming direct sums of the EJAs of the 
types mentioned above. The {\em Jordan-von Neumann-Wigner Classification Theorem} (see \cite{FK} Chapter IV) provides a converse: 
every euclidean Jordan algebra is a direct sum of {\em simple} EJAs, each of which is isomorphic to a spin factor $V_n$, or to the self-adjoint part of a matrix algebra $M_n(\K)$ where $\K$ is one 
of the classical division rings $\R, \C$ or $\Q$, 
{\em or}, if $n = 3$, to the {\em Octonions}, $\O$. This last example, which is not embeddable into the 
self-adjoint part of a complex matrix 
algebra, is called the {\em exceptional Jordan algebra}, or the {\em Albert algebra}. A Jordan algebra that 
{\em is} embeddable in $M_n(\C)_{\sa}$ for some $n$, is said to be {\em special}. It follows from the classification 
theorem that any EJA decomposes as a direct sum $A_{\sp} \oplus A_{\ex}$ where $A_{\sp}$ is special and 
$A_{\ex}$ is a direct sum of copies of the exceptional Jordan algebra.\\[0.2cm]
\noindent{\bf Projections and the Spectral Theorem} A {\em projection}
in an EJA $A$ is an element $p \in A$ with $p^2 = p$. If $p, q$ are
projections with $p \jProd q = 0$, we say that $p$ and $q$ are
orthogonal. In this case, $p + q$ is another projection. A projection
not representable as a sum of other projections is said to be {\em
  minimal} or {\em primitive}. A {\em Jordan frame} is a set $E
\subseteq A$ of pairwise orthogonal minimal projections that sum to
the Jordan unit.  The {\em Spectral Theorem}
(cf. e.g. \cite{FK}, Theorem III.1.1) for EJAs asserts that every element $a
\in A$ can be expanded as a linear combination $a = \sum_{x \in E} t_x
x$ where $E$ is some Jordan frame.\\[0.2cm]
\noindent One can show that all Jordan frames for a given Euclidean Jordan algebra $A$ have the same number of elements. This number is called the {\em rank} of $A$. By the Classification Theorem, all simple Jordan algebras having rank $4$ or higher are special.\\[0.2cm]
{\bf Order Structure} Any EJA $A$ is at the same time an ordered real vector space, 
with positive cone $A_+ = \{ a^2 | a \in A\}$; for $a, b \in A$, $a \leq b$ iff 
$b - a \in A_+$. This allows us to interpret $A$ as a probabilistic model: 
an {\em effect} (measurement-outcome) in $A$ is an element $a \in A_+$ with $a \leq u$. 
A {\em state} on $A$ is a positive linear mapping $\alpha : A \rightarrow \R$ with $\alpha(u) = 1$. 
If $a$ is an effect, we interpret $\alpha(a)$ as the probability that $a$ will 
be observed (if tested) in the state $\alpha$.\\[0.2cm] 
\noindent The cone $A_+$ is {\em self-dual} with respect to the given inner product on $A$: an element $a \in A$ belongs to $A_+$ iff $\langle a | b \rangle \geq 0$ for all $b \in A_+$.
Every state $\alpha$ then corresponds to a unique $b \in A_+$ with $\alpha(a) = \langle a | b \rangle$.\\[0.2cm]
\noindent{\em Remark:} Besides being self-dual, the cone $A_+$ is {\em homogeneous}: any element of the {\em interior} 
of $A_+$ can be obtained from any other by an order-automorphism of $A$, that is, a linear automorphism 
$\phi : A \rightarrow A$ with $\phi(A_+) = A_+$. The {\em Koecher-Vinberg Theorem} 
(\cite{Koecher, Vinberg}; see \cite{FK} for a relatively accessible proof based on 
the one in \cite{Satake})  identifies EJAs as precisely the finite-dimensional ordered linear spaces having homogeneous, self-dual positive cones. This fact underwrites the derivations in several of the papers 
cited above \cite{Wilce09, Wilce11, Wilce12}.\footnote{A different characterization of EJAs, in terms of projections 
associated with faces of the state space, is invoked in \cite{BMU}.}\\[0.2cm]
\noindent {\bf Reversible and universally reversible EJAs} A Jordan subalgebra of $M_{\sa}$, where $M$ is a
complex $\ast$-algebra, is 
{\em reversible} iff 
\[a_1,...,a_k \in A \ \Rightarrow \ a_1 a_2 \cdots a_k + a_k \cdots a_2 a_1 \in A,\]
where juxtaposition indicates multiplication in $M$. Note that with $k = 2$, this is just 
closure under the Jordan product on $M_{\sa}$.  An abstract EJA $A$ is 
{\em reversible} iff it has a representation as a reversible Jordan subalgebra of some 
complex $\ast$-algebra. A reversible EJA is {\em universally reversible} (UR) iff it 
has {\em only} reversible representations.\\[0.2cm] 
\noindent Universal reversibility will play a large role in what follows. Of the
four basic types of special Euclidean Jordan algebra considered above,
the only ones that are not UR are the spin factors $V_k$ with $k \geq
4$. For $k = 4$ and $k > 5$, $V_k$ is not even reversible; $V_5$ ---
equivalently, $M_2(\Q)_{\sa}$ --- has a reversible representation, but
also non-reversible ones.  Thus, if we adopt the shorthand
\[R_n = M_n(\R)_{\sa}, \ C_n = M_n(\C)_{\sa}, \ \mbox{and} \ Q_n = M_n(\Q)_{\sa},\]
we have $R_n, C_n$ UR for all $n$, and $Q_n$ UR for $n > 2$. 

\section{Composites of EJAs} 

A probabilistic theory must allow for some device for describing
composite systems. Given EJAs $A$ and $B$, understood as models for
two physical systems, we'd like to construct an EJA $AB$ that models
the two systems considered together as a single entity. 
Is there any satisfactory way to do this? If so, how much latitude does one have?\\[0.2cm]
\noindent A good candidate for such an EJA is provided by a construction due to Hanche-Olsen~\cite{HO}, which we now review.  In the sequel,
we show that, at least if no non-UR spin factors are involved, it does provide such a model.\\[0.2cm]
\noindent {\bf The universal tensor product} A {\em representation} of a Jordan algebra $A$ is a Jordan homomorphism $\pi : A \rightarrow\ M_{\sa}$, where 
$\M$ is a complex $\ast$-algebra. For any EJA $A$, there exists a (possibly trivial) 
${\ast}$-algebra $\Cu(A)$ and a representation $\psi_A : A \rightarrow \Cu(A)_{\sa}$ with the universal property 
that any representation $\pi : A \rightarrow\ M_{\sa}$, where $\M$ is a $C^{\ast}$-algebra, decomposes uniquely 
as $\pi = \tilde{\pi} \circ \psi_{A}$, $\tilde{\pi} : \Cu(A) \rightarrow \M$ a $\ast$-homomorphism. Evidently, 
$(\Cu(A), \psi_A)$ is unique up to a canonical $\ast$-isomorphism. Since $\psi^{\op} : A \rightarrow \Cu(A)^{\op}$ provides another solution to the same universal problem, 
there exists a canonical anti-automorphism $\Phi_{A}$ on $\Cu(A)$, fixing every point of $\psi_{A}(A)$.\\[0.2cm] 
\noindent We refer to $(\Cu(A),\psi_A)$ as the {\em universal representation} of $A$. 
$A$ is exceptional iff  $\Cu(A) = \{0\}$.  If $A$ has no exceptional factors, then $\psi_A$ is injective. In this case, we will routinely identify $A$ with its image $\psi_{A}(A) \leq \Cu(A)$.\\[0.2cm] 
\noindent In \cite{HO}, Hanche-Olsen defines the {\em universal} tensor product
of two special EJAs $A$ and $B$ to be the Jordan subalgebra of $\Cu(A)
\otimes \Cu(B)$ generated by $A \otimes B$. This is denoted $A
\hotimes B$. It can be shown that
\[\Cu(A \hotimes B) = \Cu(A) \hotimes \Cu(B) \ \ \mbox{and} \ \ \Phi_{A \hotimes B} = \Phi_{A} \otimes \Phi_{B}.\] 
Some further important facts about the universal tensor product are the following:\\[0.2cm]
\noindent {\bf Proposition 1} Let $A$, $B$ and $C$ denote EJAs. 
\begin{itemize}
\item[(a)] If $\phi : A \rightarrow C$, $\psi : B \rightarrow C$ are
  unital Jordan homomorphisms with operator-commuting
  ranges\footnote{Elements $x, y \in C$ {\em operator commute} iff
    $x \jProd(y \jProd z) = y \jProd (x \jProd z)$ for all $z \in C$.}, then
  there exists a unique Jordan homomorphism $A \hotimes B \rightarrow
  C$ taking $a \otimes b$ to $\phi(a) \jProd \psi(b)$ for all $a \in
  A$, $b \in B$.
\item[(b)] $A \hotimes B$ is UR unless one of the factors has a
  one-dimensional summand and the other has a representation onto a
  spin factor $V_n$ with $n \geq 4$.
\item[(c)]  If $A$ is UR, then $A \hotimes M_n(\C)_{\sa} = (\Cu(A) \otimes M_{n}(\C))_{\sa}$.
\end{itemize}
\noindent These are Propositions 5.2, 5.3 and 5.4, respectively, in \cite{HO}.\\[0.2cm] 
\noindent Note that part (b) implies that if $A$ and $B$ are irreducible and
non-trivial, $A \hotimes B$ will always be UR, hence, the fixed-point
set of $\Phi_A \otimes \Phi_B$.  Using this one can compute $A
\hotimes B$ for irreducible, universally reversible $A$ and $B$ \cite{HO}. Below, and 
for the balance of this paper, 
we use the shorthand $R_n := M_n(\R)_{\sa}$, $C_n = M_n(\C)_{\sa}$ and
$Q_n = M_n(\Q)_{\sa}$ (noting that $Q_n$ is UR only for $n > 2$):
\[
\begin{array}{c}
\begin{array}{l|ccc} 
\hotimes    & R_m & C_m & Q_m \\
 \hline 
R_n & R_{nm} & C_{nm} & Q_{nm}    \\
C_n & C_{nm} & C_{nm} \oplus C_{nm} & C_{2nm}  \\
Q_n & Q_{nm} & C_{2nm} & R_{4nm} 
\end{array} \\
\\
\mbox{Figure 2}
\end{array}
\]
For $Q_2 \hotimes Q_2$, a bit more work is required, but one can show that
$Q_2 \hotimes Q_2$ is the direct sum of four copies of $R_{16} = M_{16}(\R)_{\sa}$ (stated in \cite{HO}; details will be provided in \cite{BGW}). 
\\[0.2cm] 
\noindent {\bf General composites of EJAs} The universal tensor product is an instance of the following (as it proves, only slightly) more general scheme. Recall that an order-automorphism of an EJA $A$ is a linear bijection $\phi : A \rightarrow A$ taking $A_+$ onto itself. These form a Lie group, whose identity component we denote by $G(A)$.\\[0.2cm]
\noindent {\bf Definition 1:} A {\bf composite} of EJAs $A$ and $B$ is a pair $(AB,\pi)$ where $AB$ is an EJA and 
$\pi : A \otimes B \rightarrow AB$ is a linear mapping such that 
\begin{itemize} 
\item[(a)] If $a \in A_+$ and $b \in B_+$, then $\pi(a \otimes b) \in (AB)_+$, with 
$\pi (u \otimes u)$ the Jordan unit of $AB$; 
\item[(b)] for all states $\alpha $ on $A$, $\beta$ on $B$, there exists a state $\gamma$ on $AB$ such that 
$\gamma(\pi(a \otimes b)) = \alpha(a)\beta(b)$; 
\item[(c)] for all automorphisms $\phi \in G(A)$ and $\psi \in G(B)$, there exists a 
preferred automorphism $\phi \otimes \psi \in G(AB)$ with 
$(\phi \otimes \psi)(\pi(a \otimes b)) = \pi(\phi(a) \otimes \psi(b))$. 
Moreover, we require that 
\[(\phi_1 \otimes \psi_1) \circ (\phi_2 \otimes \psi_2) = (\phi_1 \circ \phi_2) \otimes (\psi_1 \circ \psi_2)\]
and
\[(\phi \otimes \psi)^{\dagger} = \phi^{\dagger} \otimes \psi^{\dagger}\] 
for all $\phi, \phi_i \in G(A)$ and $\psi, \psi_i \in G(B)$.
\end{itemize} 
\noindent It follows from (b) that $\pi$ is injective (if $\pi(T) = 0$, then for any states $\alpha, \beta$, 
there's a state $\gamma$ of $AB$ with $(\alpha \otimes \beta)(T) = \gamma(\pi(T)) = 0$; it follows that 
$T = 0$). Henceforth, we'll simply regard $A \otimes B$ as a subspace of $AB$.\\[0.2cm]
\noindent Condition (c) calls for further comment. The dynamics of a physical system modeled by a Euclidean Jordan algebra $A$ will naturally be represented by a one-parameter 
group $t \mapsto \phi_t$ of order automorphisms of $A$. As order-automorphisms in $G(A)$ are precisely the elements of such one-parameter groups, condition (c) is equivalent 
to the condition that, given dynamics $t \mapsto \phi_t$ and $t \mapsto \psi_t$ on $A$ and $B$, respectively, there is a preferred dynamics on $AB$ under which pure tensors 
$a \otimes b$ evolve according to $a \otimes b \mapsto \phi_t(a) \otimes \psi_t(b)$. In other words, there is a dynamics on $AB$ under which $A$ and $B$ evolve independently.\\[0.2cm] 
\noindent {\bf Theorem 1:} {\em If $A$ and $B$ are simple EJAs, then any composite 
$AB$ is special, and an ideal in $A \hotimes B$.}\\[0.2cm] 
\noindent The basic idea of the proof is to show that if $p_1,...,p_n$ is a Jordan frame in an irreducible summand of $A$, 
and $q_1,...,q_m$ is a Jordan frame in an irreducible summand of $B$, then $\{p_i \otimes q_j | i = 1,...,n, j = 1,...,m\}$ is a pairwise orthogonal set of projections in $AB$, whence, the latter has rank at least four, 
and must therefore be special. For the details, we refer to \cite{BGW}.
\\[0.2cm]
\noindent {\bf Corollary 1:} {\em If $A$ is simple and $B$ is exceptional, then no composite $AB$ exists.} \\[0.2cm]
\noindent In particular, if $B$ is the exceptional factor, there exists no composite of $B$ with itself.\\[0.2cm] 
\noindent {\bf Corollary 2:} {\em If $A \hotimes B$ is simple, then $AB = A \hotimes B$ is the only possible 
composite of $A$ and $B$.}\\[0.2cm] 
\noindent There are cases in which $A \hotimes B$ isn't simple, even where $A$
and $B$ are: namely, the cases in which $A$ and $B$ are both hermitian
parts of complex matrix algebras. From table (2), we see that if $A = C_n$ and $B = C_m$, then
$A \hotimes B = C_{nm} \oplus C_{nm}$. 
In this case, Proposition 1 gives us {\em two} choices for $AB$: either the entire direct sum above, 
or one of its isomorphic summands, i.e., the ``obvious" composite $AB = C_{nm}$. 

\section{Embedded EJAs} 

Corollary 1 above justifies restricting our attention to special EJAs. These are precisely the finite dimensional JC-algebras, \textit{i.e}.\ the ones isometrically isomorphic to norm closed Jordan subalgebras of self-adjoint linear operators on complex Hilbert spaces \cite{HO}\cite{HancheOlsen84}. In fact, it will be helpful to
consider {\em embedded} JC-algebras, that is, Jordan subalgebras of specified finite-dimensional $C^{\ast}$ algebras.\\[0.2cm]
\noindent {\bf Definition 2:} An {\em embedded JC algebra}, or EJC, is
a pair $(A,M_A)$ where $A$ is a unital Jordan subalgebra of a
unital finite-dimensional complex $\ast$-algebra $M_A$ (that is, a Jordan 
subalgebra such that the Jordan unit $u_A$ coincides with the unit of 
$M_A$).\\[0.2cm]
\noindent The notation $\M_A$ is intended to emphasise that the
embedding $A \mapsto \M_A$ is part of the structure of interest.
Given $A$, there is always a canonical choice for $\M_A$, namely the
universal enveloping $\ast$-algebra $\Cu(A)$ of $A$ \cite{HO}.\\[0.2cm]
{\bf Definition 3:} The {\em canonical product} of EJCs $(A,\M_A)$ and $(B,\M_B)$ is the pair $(A \odot B, \M_{A} \otimes \M_B)$ where $A \odot B$ is the Jordan subalgebra of $(\M_{A} \otimes \M_{B})_{\sa}$ generated by the subspace $A \otimes B$.\\[0.2cm]
\noindent Note that, as a matter of definition, $\M_{A \odot B} = \M_{A} \otimes \M_{B}$. 
If $\M_A = C^{\ast}(A)$ and $\M_B = C^{\ast}(B)$, then $A \odot B$ is the Hanche-Olsen tensor product.\\[0.2cm]  
\noindent One would like to know that $A \odot B$ is in fact a composite of $A$ and $B$ in the sense of Definition 1. 
Using a result of Upmeier \cite{Upmeier}, we can show that this is the case for {\em reversible} EJAs $A$ and $B$. (that is, real, complex and quaternionic systems, and direct sums of these).  Whether $A \odot B$ is a composite in the sense of Definition 1 when $A$ or $B$ is non-reversible spin factor remains an open question.\\[0.2cm] 
\noindent We can now form a category:\\[0.2cm] 
{\bf Definition 4:} $\EJC$ is the category consisting of EJCs $(A,\M_A)$ and completely positive (CP) maps $\phi : \M_A \rightarrow \M_B$ with $\phi(A) \subseteq B$. We refer to such maps as {\em Jordan preserving}.\\[0.2cm]  
{\bf Proposition 2:} {\em The canonical product $\odot$ is associative on $\EJC$. More precisely, the associator mapping 
\[\alpha : \M_{A} \otimes (\M_B \otimes \M_C) \rightarrow (\M_{A} \otimes \M_B) \otimes \M_C\]
takes $A \odot (B \odot C)$ to $(A \odot B) \odot C$. }\\[0.2cm] 
\noindent (Note that since the associator mapping is CP, this means that $\alpha$ is a morphism in $\EJC$.)  
The proof is somewhat lengthy, so we refer the reader to the forthcoming paper \cite{BGW}.\\[0.2cm] 
\noindent Proposition 2 suggests that $\EJC$ might be symmetric monoidal under $\odot$. There is certainly a natural 
choice for the monoidal unit, namely $I = (\R, \C)$. But the following 
example shows that tensor products of $\EJC$ morphisms are generally not morphisms:\\[0.2cm] 
{\bf Example 1:} Let $(A,\Cu(A))$ and $(B,\Cu(B))$ be simple, universally embedded EJCs, and suppose 
that $B$ is not UR (e.g., a spin factor $V_n$ with $n > 3$). 
Let $\hat{B}$ be the set of fixed points of the canonical involution $\Phi_B$. Then by Corollary 2, 
$A \odot B = A \hotimes B$, the set of fixed points of $\Phi_A \otimes \Phi_B$. In particular, 
$u_A \otimes \hat{B}$ is contained in $A \odot B$. Now let $f$ be a state on $\Cu(A)$: this is 
CP, and trivially Jordan-preserving, and so, a morphism in $\EJC$. But 
\[(f \otimes \id_{B})(u_A \otimes \hat{B}) = f(u_A)\hat{B} = \hat{B},\]
which isn't contained in $B$. So $f \otimes \id_{B}$ isn't Jordan-preserving. 

\section{Reversible and universally reversible EJCs}

It seems that the category $\EJC$ is simply too large.  We can try to obtain a better-behaved category by restricting the set of morphisms, or by restricting the set of objects, or both.\\[0.2cm] 
\noindent As a first pass, we might try this:\\[0.2cm]
\noindent {\bf Definition 5:} Let $(A,\M_A)$ and $(B,\M_B)$ be EJCs. A CP map $\phi : \M_A \rightarrow \M_B$ is 
{\em completely Jordan preserving} (CJP) iff $\phi \otimes 1_C$ takes $A \odot C$ to $B \odot C$ for 
all $(C,\M_C)$.\\[0.2cm] 
\noindent It is not hard to verify the following\\[0.2cm] 
{\bf Proposition 3:} {\em If $\phi : \M_{A} \rightarrow \M_B$ and $\psi : \M_C \rightarrow \M_D$ are CJP, then 
so is 
\[\phi \otimes \psi : \M_{A \odot C} = \M_{A} \otimes \M_{B} \rightarrow \M_{B} \otimes \M_{C} = \M_{B \odot D}.\]}\\[0.2cm]
\noindent Thus, the category of EJC algebras and CJP mappings  is symmetric monoidal.\footnote{Notice that 
scalars of this category are real numbers. It is sometimes suggested that quaternionic Hilbert spaces can't be 
accommodated in a symmetric monoidal category owing to the noncommutativity of $\Q$, as the scalars in a 
symmetric monoidal category must always be commutative. As we are representing 
quaternionic quantum systems in terms of the associated real vector spaces of hermitian operators, this 
issue doesn't arise here.}\\[0.2cm]
\noindent There are many examples of CJP mappings: Jordan homomorphisms are CJP maps. If $a \in A$, the mapping 
\[U_a : A \rightarrow A\]
given by $U_a = 2L_{a^2} - L_{a}^{2}$, where $L_{a}(b) = ab$, is also CJP.  On the other hand, 
by Example 1 above, $\CJP(A,I)$ is {\em empty} for universally embedded simple $A$!\\[0.2cm] 
\noindent So not all CP maps are CJP; for instance, states are never CJP. More
seriously, we can't interpret states as morphisms in this
category. The problem is the non-UR spin factors in $\CJP$.  If we
remove these, things are much better.\\[0.2cm]
\noindent {\bf Definition 6:} Let $\Cat$ be  a subclass of embedded EJC algebras, closed under $\odot$ and containing $I$. A linear mapping  $\phi : \M_{A} \rightarrow \M_{B}$ is CJP {\em relative to $\Cat$} iff 
$\phi_{A} \otimes \id_C$ is Jordan preserving for all $C$ {\em in} ${\cal C}$. 
$\CJP_{\cal C}$ is the category having objects elements of $\Cat$, mappings relatively CJP mappings.\\[0.2cm]
\noindent {\bf Example 2:} $\URUE$ is the class of universally reversible,
universally embedded EJC algebras.  $\URSE$ is the category of
universally reversible, standardly embedded EJC algebras, and $\RSE$
is the category of reversible, standardly embedded EJC
algebras. Equipped with relatively CJP mappings, both are symmetric
monoidal categories.\\[0.2cm]
\noindent Note that $\RSE$ consists of direct sums of real, complex and
quaternionic quantum systems. $\URSE$ and $\URUE$ contain all real and
complex quantum systems, and all quaternionic quantum systems {\em
  except} the ``quabit", i.e., the quaternionic bit $Q_2 :=
M_{2}(\Q)_{\sa}$.\\[0.2cm]
\noindent In both of the categories $\URUE$ and $\URSE$, states are
morphisms.  However, Example 1 provides us with a no-go result
  for extensions of these categories  that preserve 
the property that states are morphisms.  For $\URUE$, we
  cannot maintain this property while enlarging the class ${\cal C}$ 
to include a spin factor other
  than the real, complex, and quaternionic bits,
 while for $\URSE$, we cannot enlarge it to contain 
an even-dimensional one (it remains open whether we can include the
  additional odd-dimensional ones).\\[0.2cm]
\noindent  Not only do $\URUE$ and $\URSE$ include all states
as morphisms, but  we are
going to see that they inherit compact
closure from the category $\stalg$ of finite-dimensional, complex
$\ast$-algebras and CP maps, in which they are embedded.
It's worth taking a moment to review this compact structure.  If $\M$
is a finite-dimensional complex $\ast$-algebra, let $\Tr$ denote the
canonical trace on $\M$, regarded as acting on itself by left
multiplication (so that $\Tr(a) = \tr(L_a)$, $L_a : \M \rightarrow \M$
being $L_a(b) = ab$ for all $b \in \M$). This induces an inner product
on $\M$, given by $\langle a | b \rangle_{\M} = \Tr(ab^{\ast})$\footnote{We are following the convention 
that complex inner products are conjugate linear in the {\em second} argument.}.  Note
that this inner product is self-dualizing, i.e,. $a \in \M_+$ iff 
$\langle a | b \rangle \geq 0$ for all $b \in \M_+$.
Now let $\bar{\M}$ be the conjugate algebra, writing $\bar{a}$ for $a
\in \M$ when regarded as belonging to $\bar{\M}$ (so that $\bar{ca} =
\bar{c}~\bar{a}$ for scalars $c \in \C$ and $\bar{a} \bar{b} =
\bar{ab}$ for $a, b \in \M$).  Note that $\langle \bar{a}| \bar{b} \rangle = \langle b |
a \rangle$.  Now define
\[f_{\M} = \sum_{e \in E} e \otimes \bar{e} \in \M \otimes \bar{\M}\]
where $E$ is any orthonormal basis for $\M$ with respect to
$\IP_{\M}$. Then a computation shows  that
\[ \langle (a \otimes \bar{b}) f_{\M} | f_{\M} \rangle_{\M \otimes \bar{\M}} = \langle a | b \rangle_{\M}.\]
Since the left-hand side defines a positive linear functional on $\M
\otimes \bar{\M}$, so does the right (remembering here that pure
tensors generate $\M \otimes \bar{\M}$, as we're working in finite
dimensions).  Call this functional $\eta_{\M}$. That is,
\[\eta_{\M} : \M \otimes \bar{\M} \rightarrow \C \ \ \mbox{is given by} \ \ \eta_{\M}(a \otimes \bar{b}) = \langle a | b \rangle = \Tr(ab^{\ast})\]
and is, up to normalization, a state on $\M \otimes \bar{\M}$. A further computation now shows that 
\[\langle a \otimes \bar{b} | f_{\M} \rangle_{\M \otimes \bar{\M}} = \eta(a \otimes \bar{b}).\]
 It follows that $f_{\M}$ belongs to the positive cone of
$\M \otimes \bar{\M}$, by self-duality of the latter.  A final
computation  shows that, for any states $\alpha$ and
$\bar{\alpha}$ on $\M$ and $\bar{\M}$, respectively, and any $a \in
\M, \bar{a} \in \bar{\M}$, we have
\[(\eta_{\M} \otimes \alpha)(a \otimes f_{\bar{\M}}) = \alpha(a) \ \ \mbox{and} \ \ (\bar{\alpha} \otimes \eta_{\M})(f_{\bar{M}} \otimes \bar{a}) = \bar{\alpha}(\bar{a}).\]
Thus, $\eta_{\M}$ and $f_{\bar{\M}}$ define a compact structure on
$\stalg$, for which the dual object of $\M$ is given by $\bar{\M}$.\\[0.2cm]
\noindent {\bf Definition 7:} The {\em conjugate} of a EJC algebra $(A,\M_{A})$ is $(\bar{A},\bar{\M}_{A})$, where 
$\bar{A} = \{ \bar{a} | a \in A\}$. We write $\eta_{A}$ for $\eta_{\M_A}$ and $f_A$ for $f_{\M_A}$. 

\subsection{Universally-embedded, universally reversible EJAs}

Now consider the category $\URUE$ of universally reversible, universally embedded EJAs $A$, i.e., 
pairs $(A,\M_A)$ with $A$ UR and $\M_A = \Cu(A)$. Let $\Phi_A$ be the canonical involution on $\Cu(A)$. \\[0.2cm]
\noindent{\bf Lemma 1:} {\em Let $(A,\M_A)$ belong to $\URUE$. Then 
\begin{itemize} 
\item[(a)] $f_{A} \in A \odot \bar{A}$; 
\item[(b)] $\eta_{A} \circ (\Phi_{A} \otimes \Phi_{\bar{A}}) = \eta_{A}$. 
\end{itemize} }
\noindent {\em Proof:} (a) follows from the fact that $\Phi_A$ is unitary, so
that if $E$ is an orthonormal basis, then so is $\{\Phi_{A}(e) | e \in
E\}$.  Since $f_{A}$ is independent of the choice of orthonormal
basis, it follows that $f$ is invariant under $\Phi_{A} \otimes
\Phi_{\bar{A}}$, hence, an element of $A \hotimes \bar{A}$. Now (b)
follows from part (a) of the previous lemma. $\Box$\\[0.2cm]
\noindent Define $\gamma_A : \Cu(\bar{A}) \rightarrow \Cu(A)$ by $\gamma(\bar{a}) = \Phi_{A}(a^{\ast})$. 
This is a $\ast$-isomorphism, and intertwines $\Phi_{A}$ and $\Phi_{A}$; hence, 
$\gamma_{A} \otimes \id_{B} : \Cu(\bar{A} \hotimes B)  
\rightarrow \Cu(A \otimes B)$ intertwines 
$\Phi_{\bar{A}} \otimes \Phi_{B} = \Phi_{\bar{A} \hotimes B}$ and 
$\Phi_{A} \otimes \Phi_{B} = \Phi_{A \hotimes B}$ 
--- whence, takes $\bar{A} \hotimes B$ to $A \hotimes B$. In particular, $\gamma_{A}$ is CJP relative 
to the class of UR, universally embedded EJCs.\\[0.2cm] 
{\bf Lemma 2:} {\em Let $A$ be a universally embedded UR EJC. Then for all $\alpha \in \CJP(A,I)$, 
there exists $a \in A$ with $\alpha(b) = \langle b |a \rangle$ for all $b \in A$. }\\[0.2cm] 
\noindent {\em Proof:} Since $\alpha \in \Cu(A)^{\ast}$, there is certainly some $a \in \Cu(A)$ with 
$\alpha = |a \rangle$. We need to show that $a \in A$. Since $\alpha$ is CJP, 
\[\gamma_{A} \otimes \alpha : \Cu(\bar{A}) \otimes \Cu(A) = \Cu(\bar{A} \otimes A) \rightarrow \Cu(A)\]
is Jordan-preserving. In particular, $(\alpha \otimes \gamma_{A})(f_A) \in A$.  But 
\begin{eqnarray*} 
(\alpha \otimes \gamma_A)(f_A) & = & \sum_{e \in E}  (\alpha \otimes \gamma_{A})(e \otimes \bar{e}) \\
& = & \sum_{e \in E} \langle e | a \rangle \Phi(e^{\ast}) \\
& = & \Phi(\sum_{e \in E} \langle e | a \rangle e^{\ast}) \\
& = & \Phi((\sum_{e \in E} \langle a | e \rangle e )^{\ast}) = \Phi(a^{\ast}) = \gamma_{A}(\bar{a}).
\end{eqnarray*}
Hence, $\gamma_{A}(\bar{a}) \in A$, whence, $\bar{a} \in \bar{A}$, whence, $a \in A$. (Alternatively: 
$\Phi_{A}(a^{\ast}) \in A$ implies $a^{\ast} \in A$, whence, $a$ is self-adjoint, whence, $a \in A$.) $\Box$\\[0.2cm]
\noindent It follows that $\eta_{A}$ and $f_A$ belong, as morphisms, to $\URUE$. Hence, $\URUE$ inherits the 
compact structure from $\stalg$, as promised.\\[0.2cm] 
\noindent The same holds for $\URSE$. Specifically, we want to show that $f_{A}$ belongs to $A \odot \bar{A}$ whenever 
$A$ is a standardly embedded UR EJC.\\[0.2cm] 
\noindent Suppose that $E$ is an orthonormal basis for $\M_{A}$: 
then so is
 $\{ e^{\ast} | e \in E\}$; thus, since $f_{A}$ is independent of the choice of basis, we have 
 \[f_{A}^{\ast} = \sum_{e \in E} e^{\ast} \otimes \bar{e}^{\ast} = \sum_{e^{\ast} \in E^{\ast}} e^{\ast} \otimes \bar{e^{\ast}} = f_{A}.\]
 Thus, if $A \odot \hat{A}$ is the self-adjoint part of $\M_{A} \otimes \bar{\M}_{A}$, then $f_{A} \in A \otimes \bar{A}$. This covers the case where $A = C_n$. We also have, by the results above, that 
 $f_{A} \in A \odot \bar{A}$ whenever the latter equals $A \hotimes \bar{A}$. 
 This covers $A = R_n$ and $A = Q_n$ for $n > 2$.\\[0.2cm]
\noindent In fact, we can do a bit better. If $\M$ and $\N$ are finite-dimensional $\ast$-algebras and $\phi : \M \rightarrow \N$ is a linear mapping, let $\phi^{\dagger}$ denote the adjoint 
of $\phi$ with respect to the natural trace inner products on $\M$ and $\N$. It is not difficult to show 
that, for any $\M$ in $\stalg$, $f_{\M}^{\dagger} = \eta_{\M}$ and vice versa; indeed, $\stalg$ is dagger-compact.\\[0.2cm] 
\noindent {\bf Definition 8:} Let $(A,\M_A)$ and $(B,\M_B)$ be EJCs. A linear mapping $\phi : \M_A \rightarrow \M_B$ is 
{\em $\dagger$-CJP} iff both $\phi$ and $\phi^{\dagger}$ are CJP. If $\Cat$ is a category of EJCs and 
CJP mappings, we write $\Cat^{\dagger}$ for the category having the same objects, but with morphisms restricted 
to $\dagger$-CJP mappings in $\Cat$.\\[0.2cm] 
\noindent If $A$ belongs to $\URUE$ or $\URSE$, then $f_{A}$ and $\eta_{A}$ are both CJP and, hence, are both $\dagger$-CJP with respect to the indicated category. Hence,\\[0.2cm] 
\noindent {\bf Theorem 2:} {\em The categories $\URUE^{\dagger}$ and $\URSE^{\dagger}$ are dagger-compact.}

\section{Conclusion} 

\noindent We have found two theories --- the categories $\URSE$ and $\URUE$ of 
universally reversible euclidean Jordan algebras, standardly and universally embedded --- that, in slightly different ways, unify 
finite-dimensional real, complex and (almost all of) quaternionic quantum mechanics.  By virtue of being compact closed, both theories continue to enjoy many of the information-processing properties of standard complex QM, e.g., the existence of conclusive teleportation and entanglement-swapping protocols \cite{Abramsky-Coecke}.\\[0.2cm] 
One property of standard QM that some of our composites do \emph{not} share is the possibility of \emph{local tomography}:  a state $\omega$ on $A
\odot B$ is not generally determined by the joint probability
assignment $a,b \mapsto \omega(a \otimes b)$, where $a$ and $b$ are
effects of $A$ and $B$, respectively. Another way to put it is that $A
\odot B$ is generally much larger than the vector-space tensor product
$A \otimes B$. As local tomography is well known to separate complex
QM from its real and quaternionic variants, this is hardly surprising.
Perhaps more surprising is that some of our composites---those
  involving quaternionic factors---also exhibit supermultiplicativity of
  the maximal number of distinguishable states, and, relatedly,
  violate another property sometimes assumed as a basic desideratum
  for composites: that products of pure states are pure.\\[0.2cm]
\noindent Neither theory includes the quabit, $Q_2$. Example 1 shows that the quabit can't be added to $\URUE$ without a violation of compact
closure. On the other hand, if $f_{Q_2}$ belongs to the canonical
composite $Q_2 \odot Q_2$, then the slightly larger category $\RSE$,
which consists of {\em all} finite-dimensional real, complex and
quaternionic quantum systems, will be compact closed (indeed, dagger
compact).\footnote{Since this note was accepted for presentation at
QPL XII 
we have settled this question in the affirmative. The details will 
be given in \cite{BGW}.}\\[0.2cm]
\noindent The categories $\URSE$ and $\URUE$ contain interesting compact closed
subcategories.  In particular, real and quaternionic quantum systems (less the quabit), taken
together, form a sub-theory, closed under composites. We {\em
  conjecture} that this is exactly what one gets by applying the CPM
construction to Baez's (implicit) category of pairs $(\H,J)$, $\H$ a
finite-dimensional Hilbert space and $J$ an anti-unitary with $J^2 =
\pm 1$ --- and, again, excluding the quabit. Should $\RSE$ prove to be
compact closed, we could entertain the stronger conjecture that this
is exactly what one obtains by applying CPM to Baez's category.\\[0.2cm]
\noindent Complex quantum systems also form a monoidal subcategory of $\URSE$,
which we might call $\CQM$: indeed, one that functions as an ``ideal",
in that if $A \in \URSE$ and $B \in \CQM$, then $A \odot B \in \CQM$
as well.  This is provocative, as it suggests that a universe
initially consisting of many systems of all three types, would
eventually evolve into one in which complex systems greatly
predominate.\\[0.2cm]
\noindent The category $\URUE$ is somewhat mysterious. Like $\URSE$, this
encompasses real, complex and quaternionic quantum systems, excepting
the quabit. In this theory, the composite of {\em complex} quantum
systems comes with an extra classical bit --- equivalently, a
$\{0,1\}$-valued superselection rule. This functions to make the 
transpose operation --- which is a Jordan automorphism of $\M_n(\C)_{\sa}$, 
but an antiautomorphism of $\M_n(\C)$ --- count as a morphism. The
precise physical significance of this is a subject for further study.
\section*{Acknowledgements}
We thank Cozmin Ududec for valuable discussions. Research at Perimeter Institute is supported by the Government of Canada through Industry Canada and by the Province of Ontario through the Ministry of Research \& \mbox{Innovation}. MAG was supported by an NSERC Alexander Graham Bell Canada Graduate Scholarship. AW was supported by a grant from the FQXi foundation.

\nocite{*}
\bibliographystyle{eptcs}
\bibliography{snqtARXV3Ref}
\end{document}